\documentclass[runningheads]{llncs}
\usepackage{amsmath}
\usepackage{amssymb}
\usepackage{graphicx}
\usepackage{subfigure}
\usepackage{color}
\usepackage{url}

\begin{document}

\mainmatter  

\title{Discrete-time Quantum Walk on the Cayley Graph of the Dihedral Group}

\titlerunning{Discrete-time Quantum Walk on the Cayley Graph of the Dihedral Group}
%
%
\author{Wenjing Dai\footnote{Electronic address: jing@nuaa.edu.cn.}{$^{,1}$}, Jiabin Yuan\footnote{Email: jbyuan@nuaa.edu.cn.}{$^{,1}$} \and Dan Li\footnote{Email: lidansusu007@163.com.}{$^{,1}$}}
\authorrunning{Wenjing Dai, Jiabin Yuan and Dan Li}
\institute{{$^{1}$}College of Computer Science and Technology, Nanjing University of Aeronautics and Astronautics, Nanjing, 211106, China\\
}


\toctitle{Lecture Notes in Computer Science}
\tocauthor{Authors' Instructions}
\maketitle

\begin{abstract}
The finite dihedral group generated by one rotation and one flip is the simplest case of the non-abelian group. Cayley graphs are diagrammatic counterparts of groups. In this paper, much attention is given to the Cayley graph of the dihedral group. Considering the characteristics of the elements in the dihedral group, we conduct the model of discrete-time quantum walk on the Cayley graph of the dihedral group by special coding mode. This construction makes Fourier transformation can be used to carry out spectral analysis of the dihedral quantum walk, i.e. the non-abelian case. Furthermore, the relation between quantum walk without memory on the Cayley graph of the dihedral group and quantum walk with memory on a cycle is discussed, so that we can explore the potential of quantum walks without and with memory. Here, the numerical simulation is carried out to verify the theoretical analysis results and other properties of the proposed model are further studied.
\end{abstract}

\section{Introduction}

Quantum computer is seen as a new kind of computer, which has outstanding computing power. It has been proven that Turing un-computability problem, such as the halting problem, also can not be calculated in quantum computing. Therefore, quantum computing gets no essential breakthrough from the computability point of view. However, the power of quantum computing is likely to be stronger than classical Turing machine from the view of computational complexity. The best evidence is that Shor's algorithm and Grover's algorithm have superior computing power over their corresponding classical algorithms. In 1994, Shor proposed a quantum algorithm for solving integer factoring and discrete logarithm problem. The algorithm provides a striking exponential speedup over all known classical algorithms \cite{shor1994algorithms}. In 1996, Grover proposed the quantum search algorithm on an unordered database, which provides quadratic speedup over many classical heuristic search algorithms \cite{grover1996fast}.

There are significant progress in designing novel quantum algorithms in recent years, such as quantum walks \cite{ambainis2008quantum,liu2018hypergraphs} and quantum machine learning \cite{lloyd2010quantum,duan2017smm}. Quantum walks are the quantum version of classical random walks. Quantum walks provide a general framework for designing quantum algorithms. Given a graph structure, quantum walks simulate the behavior of particles on the graph based on the quantum mechanical properties, which is different from classical random walks. There are two classes of quantum walks, that is, the discrete-time and continuous-time quantum walks. We focus on the discrete-time model. The two models of quantum walks on the line and general graphs have attached much attention.

Aharonov proposed the concept of discrete quantum walks with quantum coin tossing in one-dimensional Hilbert space and successfully introduced random walks into the field of quantum computing \cite{aharonov1993quantum}. Nayak and Vishwanath analyzed in detail the dynamics of the one-dimensional Hadamard walk. They take the Sch\"{o}dinger approach to determine the asymptotic form of the probability distribution induced by the line by observing the position of a particle doing the walk \cite{nayak2000quantum}. Ambainis et al. gave two general ideas for analyzing quantum random walks, i.e. the path integral approach and the Sch\"{o}dinger approach, to study intensively the Hadamard walk with one and two absorption boundaries \cite{ambainis2001one}. In contrast with the central limit theorem for classical random walks, Konno showed a new type of weak limit theorems for the one-dimensional lattice \cite{konno2002quantum,konno2005limit}. After that, Konno had a deep study on the probability properties, i.e. the localization theorem and the balance of probability distribution of discrete quantum walks in one dimension \cite{konno2002limit,konno2003absorption,konno2004symmetry,konno2005path,konno2009limit,konno2009one,konno2010local,konno2011local,konno2013limit}.

Tregenna et al. studied discrete quantum walks on regular lattices, in one and two spatial dimensions. They used the bias in the coin operator, the phases in the coin operator and the initial state of the coin to control the evolution of the quantum walk \cite{tregenna2003control}. Brun et al. first analyzed the one-dimensional quantum walk driven by many coins \cite{brun2003quantum}.

Since various quantum algorithms based on quantum walks technology focused on quantum walks on graphs, a review is essential. Aharonov et al. studied several properties of quantum walks on undirected graphs. They gave definitions of mixing time, filling time, dispersion time and set the ground for this theory. Quantum walk on a cycle was taken as an example to use the eigenvalues and eigenvectors of the unitary operator of the walk to discuss the limiting distribution and the mixing time of the quantum walk \cite{aharonov2001quantum}. Henceforth, various researchers studied the properties of quantum walks on cycles \cite{bednarska2003quantum,bednarska2004quantum,inui2005flut,banerjee2008symmetry}.

A cycle with $n$ nodes is 2-regular graph, which can be viewed as the Cayley graph of an Abelian group. Cayley graphs are diagrammatic counterparts of groups. Cayley graphs are convenient means to study quantum walks exploiting the group-theoretical machinery. In addition, quantum walk on the hypercube receives a great deal of attention, which is also quantum walk on the Cayley graph of the Abelian group. Moore and Russell studied the discrete-time and continuous-time quantum walks on the hypercube. They also proved the mean inversion operator, i.e. Grover's diffusion operator, could enable the walk to achieve the fastest diffusion speed \cite{moore2002quantum}. Kempe provided the basic framework for quantum hitting time and gave two alternative definitions. His work laid a foundation for quantum walks on general graphs. It is also pointed out that quantum walk on the hypercube has exponential speedup over its classical counterpart in Ref. \cite{kempe2005quantum}. Shenvi et al. proposed a quantum search algorithm based on quantum walk on the hypercube, i.e. the SKW algorithm. The SKW algorithm provided a quadratic speedup over its classical algorithm \cite{shenvi2003quantum}.

These research findings can be viewed as quantum walks on infinite or finite Abelian groups. However, quantum walks on non-abelian groups are more suitable for modeling complicated situations than quantum walks on Abelian groups. The dihedral group is one of the simplest non-abelian groups. This paper focuses on discrete quantum walks on dihedral groups. For simplicity, we pay much attention to discrete-time quantum walk on the Cayley graph of the dihedral group.

The rest of the paper is organized as follows. Necessary preliminaries are listed in Section 2. The model of discrete-time quantum walk on the Cayley graph of the dihedral group is introduced in Section 3. The spectral analysis of its evolution operator using Fourier transformation is given in Section 4. The relation between quantum walk without memory on the Cayley graph of the dihedral group $D_N$ and quantum walk with memory on an $N$-cycle is discussed in Section 5. The numerical simulation of the introduced model is provided in Section 6. Finally, a short summary of the obtained results is given in Section 7.

\section{Preliminaries}

Several fundamental definitions and notions are introduced in this part \cite{serre2012linear,qiu2011group,gary2007graph,venegas2012review,portugal2013search}.

\begin{definition}
\textbf{(Dihedral group)}
The dihedral group ${D_N}$ is a symmetric group of $N$-gons ($N \ge 3$). Let $\sigma$ denote the rotation of the regular $N$-polygon by an angle of ${{2\pi } \mathord{\left/ {\vphantom {{2\pi } N}} \right. \kern-\nulldelimiterspace} N}$ degrees. Let $\tau $ denote the flip of the regular $N$-polygon around an axis of symmetry. Then ${D_N} =  < \sigma ,\tau |{\sigma ^N} = {\tau ^2} = {1_{{D_N}}},\tau \sigma \tau  = {\sigma ^{ - 1}} > $, where $\sigma$ and $\tau$ are two generators, and $1_{{D_N}}$ is the identity element of $D_N$.
\end{definition}

\paragraph{\textbf{Remark}}

The dihedral group is denoted by ${D_N}$ or ${D_{2N}}$ in different documents. Traditionally, geometers like to use ${D_N}$ to emphasize the number of sides of a regular polygon, while algebraists like to use ${D_{2N}}$ to emphasize the order of the symmetric group of regular polygons.

\begin{definition}
\textbf{(Semi-direct product)}
Let $G$ be a group and $N \triangleleft G$ (normal subgroup), $H < G$ (proper subgroup). If $G = NH$ and $N \cap H = \left\{ e \right\}$, where $e$ is the identity of $G$, then $G$ is called a semi-direct product of $N$ and $H$, denoted by ${G} \cong {N}\rtimes{H}$.
\end{definition}

The dihedral group ${D_N}$ is isomorphic to the semi-direct product of ${Z_N}$ and ${Z_2}$, denoted by ${D_N} \cong {Z_N}\rtimes{Z_2}$. Each element of the dihedral group can be expressed as ${\tau ^s}{\sigma ^t}$, written as a pair $(s, t)$, where $s \in {Z_2},t \in {Z_N}$ and ${Z_N} = \left\{ {0,1, \cdots ,N - 1} \right\}$. If $s = 0$, $(s, t)$ is called a rotation of the dihedral group. If $s = 1$, $(s, t)$ is called a flip of the dihedral group.

\begin{definition}
\textbf{(Cayley graph)}
Let $G$ be a finite group, and let $H = \left\{ {{h_1}, \cdots ,{h_k}} \right\}$ be a generating set for $G$. The Cayley graph of $G$ with respect to $H$ has a vertex for every element of $G$, with an oriented edge from $g$ to $gh$, where $\forall g \in G$ and $\forall h \in H$.
\end{definition}

\begin{figure}
\centering
\subfigure[]{
\includegraphics[height=2.8cm]{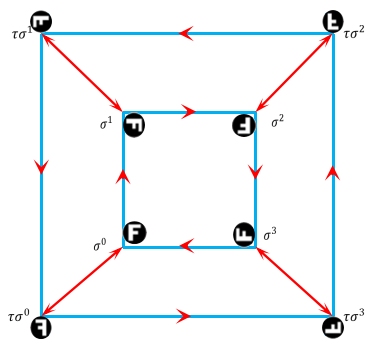}
}
\subfigure[]{
\includegraphics[height=3.0cm]{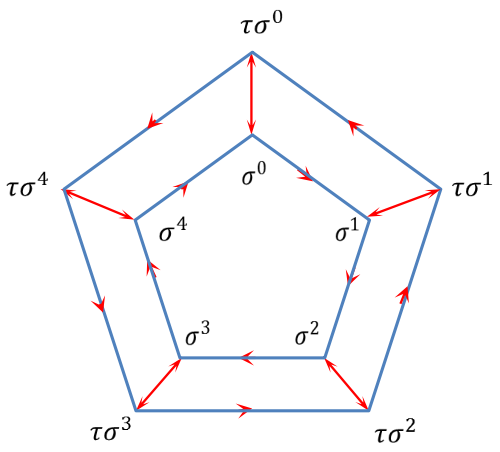}
}
\subfigure[]{
\includegraphics[height=2.9cm]{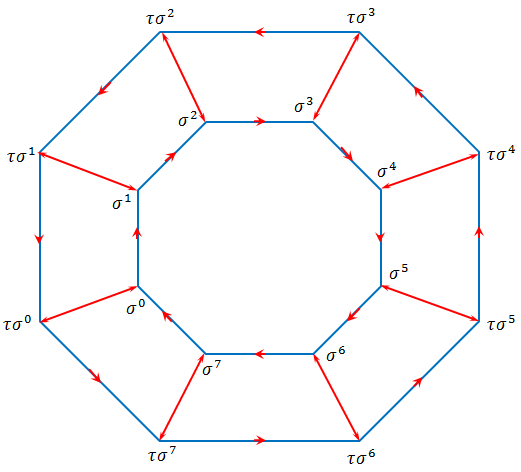}
}
\caption{Subgraph (a), (b) and (c) show the Cayley graph of the dihedral group ${D_4} = \{ e,{\rm{ }}\sigma ,{\rm{ }}{\sigma ^2},{\rm{ }}{\sigma ^3},\tau ,{\rm{ }}\tau \sigma ,{\rm{ }}\tau {\sigma ^2},{\rm{ }}\tau {\sigma ^3}\} $, ${D_5}$ and ${D_8}$, respectively.}
\label{fig:1}
\end{figure}

Several examples of the Cayley graph of the dihedral group are given in Fig.\ref{fig:1}.

Recall that a walker doing a random walk on the dihedral group chooses, at every time step, a random direction to move in, and moves to the site adjacent to it in that direction by assigning probability ${1 \mathord{\left/{\vphantom {1 d}} \right.\kern-\nulldelimiterspace} d}$ to all edges leaving each vertex $v$ of degree $d$.

\begin{definition}
\textbf{(Classical random walks on dihedral groups)}
Let $G = {D_N}$ for an integer $N \ge 3$. Define a rotation $R = {\sigma ^t}$ with $0<t<N$ and a flip $F = \tau {\sigma ^s}$ with $0 \le s < N$. Define
\begin{equation}
P\left( g \right) = \left\{ {\begin{array}{*{20}{c}}
{{1 \mathord{\left/
 {\vphantom {1 2}} \right.
 \kern-\nulldelimiterspace} 2},if\begin{array}{*{20}{c}}
{}
\end{array}g = {\sigma ^t}}\\
{{1 \mathord{\left/
 {\vphantom {1 2}} \right.
 \kern-\nulldelimiterspace} 2},if\begin{array}{*{20}{c}}
{}
\end{array}g = \tau {\sigma ^s}}\\
{0,otherwise}
\end{array}} \right.,  \label{1}
\end{equation}
where $g \in G$ and $\left| {t - s} \right|\left( {\bmod N} \right) = 1$.
\end{definition}
\begin{definition}

\textbf{(Coined quantum walks)}
Let $G(V,E)$ be a graph. Let ${H_V}$ be the Hilbert space spanned by the vertex states $\left| v \right\rangle $, where $v \in V$. The number of vertices in $G(V,E)$ is denoted by $\left| V \right|$. Assume that $G(V,E)$ is $d$-regular. Let ${H_C}$ be an auxiliary Hilbert space of dimension $d$ spanned by the states $\left| 1 \right\rangle $ through $\left| d \right\rangle $. Let $\hat C$ be the coin-tossing operator on ${H_C}$. Label each directed edge with a number between 1 and $d$, such that for each $a$, the directed edges labeled $a$ form a permutation. For Cayley graphs, the labeling of a directed edge is simply the generator associated with the edge. Define a shift operator $\hat S$ on ${H_C} \otimes {H_V}$ such that $\hat S\left| {c,v} \right\rangle  = \left| {c,u} \right\rangle $, where $u$ is the  $c$th neighbor of $v$. Note that since the edge labeling is a permutation, $S$ is unitary. One step of the quantum walk is given by $\hat U = \hat S \cdot \left( {\hat C \otimes \hat I_V} \right)$.
\end{definition}

The main components of a coined quantum walk are a walker, a coin, evolution operators for both walker and coin, and a set of observables \cite{venegas2012review}. Randomness is introduced by measuring quantum systems.

\section{Model}

In this part, we consider the Cayley graph of the finite dihedral group $D_N$ which is a 2-regular graph. The Hilbert space of the walk is $H_C^2 \otimes H_V^{2N}$.

\subsection{Feasibility Analysis: Reversibility}

Aharonov et al. defined a coined quantum walk on a $d$-regular undirected graph \cite{aharonov2001quantum}. Montanaro studied quantum walks on directed graphs and showed that reversibility is a necessary and sufficient condition for a directed graph \cite{montanaro2005directed}, using a similar definition to Ref. \cite{aharonov2001quantum}. Because the Cayley graph of the dihedral group is a directed graph, as a first step, feasibility analysis is essential.

\begin{theorem}
A discrete-time quantum walk can be defined on a finite graph $G(V,E)$ if and only if $G(V,E)$ is reversible.
\end{theorem}

According to Ref. \cite{montanaro2005directed}, all Cayley graphs are regular, and hence are reversible. Because the Cayley graph of the dihedral group is 2-regular and reversible, discrete-time quantum walk on the Cayley graph of the dihedral group can be defined.

\subsection{Extended Coined Model and Evolution}

\begin{figure}
\centering
\includegraphics[height=3.8cm]{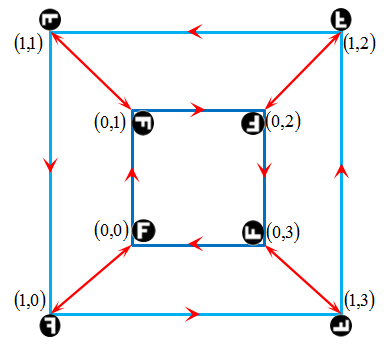}
\caption{Good encoding of the Cayley graph of the dihedral group $D_4$.}
\label{fig:2}
\end{figure}

In this part, the model of discrete-time quantum walk on the Cayley graph of the dihedral group is defined. Each element of the dihedral group is denoted by a pair $(s, t)$. Thus vertices of the Cayley graph of the dihedral group can be encoded as a pair $(s, t)$ and the pair $(s, t)$ corresponds to two registers $\left| s \right\rangle \left| t \right\rangle $. The vertex set is denoted by
\begin{equation}
V = \left\{ {\overbrace {\left| 0 \right\rangle \left| 0 \right\rangle ,\left| 0 \right\rangle \left| 1 \right\rangle , \cdots ,\left| 0 \right\rangle \left| {N - 1} \right\rangle }^{rotation},\underbrace {\left| 1 \right\rangle \left| 0 \right\rangle ,\left| 1 \right\rangle \left| 1 \right\rangle , \cdots ,\left| 1 \right\rangle \left| {N - 1} \right\rangle }_{flip}} \right\}. \label{2}
\end{equation}
The example $D_4$ is shown in Fig.\ref{fig:2}.

Considering the dihedral group generated by one rotation and one flip, each vertex has two directions, i.e. $C \in SU\left( 2 \right)$, where $SU\left( 2 \right)$ is the group of ${\rm{2}} \times {\rm{2}}$ unitary matrices of determinant 1. Among coin operators, the Hadamard operator is extensively employed. Thus the Hadamard operator is chosen as the coin tossing operator, which is as follows
\begin{equation}
\hat H = \frac{1}{{\sqrt 2 }}\left( {{{\left| 0 \right\rangle }_C}\langle 0| + {{\left| 0 \right\rangle }_C}\langle 1| + {{\left| 1 \right\rangle }_C}\langle 0| - {{\left| 1 \right\rangle }_C}\langle 1|} \right) .\label{3}
\end{equation}
Edges of the Cayley graph of the dihedral group can be encoded as $\left| {direction} \right\rangle \left| {vertex} \right\rangle $ by an auxiliary register attached to the registers of the vertex set, i.e.
\begin{equation}
E = \left\{ {\overbrace {\left| R \right\rangle \left| 0 \right\rangle \left| 0 \right\rangle , \cdots ,\left| R \right\rangle \left| 1 \right\rangle \left| {N - 1} \right\rangle }^{\left| R \right\rangle },\underbrace {\left| F \right\rangle \left| 0 \right\rangle \left| 0 \right\rangle , \cdots ,\left| F \right\rangle \left| 1 \right\rangle \left| {N - 1} \right\rangle }_{\left| F \right\rangle }} \right\}.  \label{4}
\end{equation}
The shift operator is defined by
\begin{equation}
\begin{array}{*{20}{l}}
{\hat S{{\left| R \right\rangle }_C}\left| 0 \right\rangle \left| j \right\rangle  = {{\left| R \right\rangle }_C}\left| 0 \right\rangle \left| {j + 1} \right\rangle }\\
{\hat S{{\left| F \right\rangle }_C}\left| 0 \right\rangle \left| j \right\rangle  = {{\left| F \right\rangle }_C}\left| 1 \right\rangle \left| j \right\rangle }\\
{\hat S{{\left| R \right\rangle }_C}\left| 1 \right\rangle \left| j \right\rangle  = {{\left| R \right\rangle }_C}\left| 1 \right\rangle \left| {j - 1} \right\rangle }\\
{\hat S{{\left| F \right\rangle }_C}\left| 1 \right\rangle \left| j \right\rangle  = {{\left| F \right\rangle }_C}\left| 0 \right\rangle \left| j \right\rangle }\\
{j \in {Z_N}}
\end{array}. \label{5}
\end{equation}

The shift operator allows the walker to go one step rotation if the accompanying coin state is ${\left| R \right\rangle _C}$, or one step flip if the accompanying coin state is the other basis state ${\left| F \right\rangle _C}$. A suitable shift operator of discrete-time quantum walk on the Cayley graph of the dihedral group has the following form
\begin{equation}
\begin{array}{*{20}{l}}
{\begin{array}{*{20}{l}}
{\hat S = \left| R \right\rangle \left. {\langle R} \right| \otimes \left| 0 \right\rangle \left. {\langle 0} \right| \otimes \sum\limits_j {\left| {j + 1\left( {{\mkern 1mu} \,\bmod \,{\mkern 1mu} N} \right)} \right\rangle \left. {\langle j} \right|} }\\
{ + \left| R \right\rangle \left. {\langle R} \right| \otimes \left| 1 \right\rangle \left. {\langle 1} \right| \otimes \sum\limits_j {\left| {j - 1\left( {{\mkern 1mu} \,\bmod \,{\mkern 1mu} N} \right)} \right\rangle \left. {\langle j} \right|} }\\
{ + \left| F \right\rangle \left. {\langle F} \right| \otimes \left| 0 \right\rangle \langle 1| \otimes \sum\limits_j {\left| {j\left( {{\mkern 1mu} \,\bmod \,{\mkern 1mu} N} \right)} \right\rangle \left. {\langle j} \right|} }
\end{array}}\\
{ + \left| F \right\rangle \left. {\langle F} \right| \otimes \left| 1 \right\rangle \langle 0| \otimes \sum\limits_j {\left| {j\left( {{\mkern 1mu} \,\bmod \,{\mkern 1mu} N} \right)} \right\rangle \langle j|} }
\end{array}. \label{6}
\end{equation}
The shift operator $\hat S $ controls the walker to perform rotation one unit if the coin state is in the ``$R$" state or to perform flip one unit if the coin state is in the ``$F$" state. If the walker is in some superposition of ``$R$" and ``$F$", the terms evolve accordingly. $\hat C$ acts on the coin state space and $\hat S$ acts on both the vertex state space and coin space together. Consequently, the operator for each step on the total Hilbert space is unitary and has the form:
\begin{equation}
\hat U = \hat S \cdot \left( {\hat C \otimes {{\hat I}_V}} \right), \label{7}
\end{equation}
where ${{{\hat I}_V}}$ is the identity operator on the vertex space.

\section{Spectral Analysis}

The vertex state space on which the walker moves is the vertex set of the Cayley graph of the dihedral group $D_N$. The spatial part has associated a $2N$-dimensional Hilbert space $H_V^{2N}$ with computational basis $\left\{ {\left| j \right\rangle ,0 \le j \le N - 1} \right\}$, where the values of $j$ correspond to the labels $t$ of the vertices denoted by a pair $(s, t)$. The coin state space is two-dimensional, because the walker can move rotation or flip. Thus, the Hilbert space associated with the quantum walk is ${H_C^2} \otimes {H_V^{2N}}$, with computational basis $\left\{ {\left| {c,s,j} \right\rangle ,0 \le c \le 1 ,0 \le s \le 1,0 \le j \le N - 1} \right\}$. $c = 0$ represents rotation transformation and $c = 1$ represents flip transformation. $s = 0$ represents the rotation elements of the dihedral group and $s = 1$ represents the flip elements of the dihedral group.

The shift operator constructed by Eq. (\ref{6}) is simplified as
\begin{equation}
\begin{array}{*{20}{l}}
{\hat S\left| {0,s,j} \right\rangle  = \left| {0,s,j + {{\left( { - 1} \right)}^s}} \right\rangle }\\
{\hat S\left| {1,s,j} \right\rangle  = \left| {1,\bar s,j} \right\rangle }
\end{array}, \label{8}
\end{equation}
where $\bar{s}=1-s$. Arithmetic operations with variable $j$ are performed modulo $N$. According to the coding scheme above, the walk operator is defined as follows
\begin{equation}
U = S\left( {C \otimes {{\rm I}_{2N}}} \right) = S\left( {C \otimes {{\rm I}_2} \otimes {{\rm I}_N}} \right), \label{9}
\end{equation}
where $C$ is a coin matrix.

Nayak and Vishwanath \cite{nayak2000quantum}, followed by Ambainis et al. \cite{ambainis2001one}, used the discrete time Fourier transform and methods of complex analysis to analyze the properties of discrete quantum walk on the line. The walk starts in some initial state $\left| {{\psi _0}} \right\rangle $. Following Refs. \cite{nayak2000quantum,ambainis2001one}, the quantum state is changed according to the formula
\begin{equation}
\left| {{\psi _n}} \right\rangle  = {U^n}\left| {{\psi _0}} \right\rangle  \label{11}
\end{equation}
or as a recursive relation
\begin{equation}
\left| {{\psi _{n + 1}}} \right\rangle  = U\left| {{\psi _n}} \right\rangle. \label{12}
\end{equation}

For the Hadamard walk, the evolution is given as follows
\begin{equation}
U = S\left( {H \otimes {{\rm I}_{2N}}} \right) = S\left( {H \otimes {{\rm I}_2} \otimes {{\rm I}_N}} \right), \label{13}
\end{equation}
where $H$ is Hadamard matrix
\begin{equation}
H = \frac{1}{{\sqrt 2 }}\left( {\begin{array}{*{20}{c}}
1&1\\
1&{ - 1}
\end{array}} \right). \label{10}
\end{equation}

The coin part in Eq. (\ref{13}) can be written in the matrix notation as follows
\begin{equation}
{H \otimes {\rm I}_{2N}} = \frac{1}{{\sqrt 2 }}\left( {\begin{array}{*{20}{c}}
{{{\rm I}_{2N}}}&{{{\rm I}_{2N}}}\\
{{{\rm I}_{2N}}}&{ - {{\rm I}_{2N}}}
\end{array}} \right). \label{14}
\end{equation}

Here we calculate amplitudes for a quantum walk on the Cayley graph of the dihedral group. In this case, the vectors of amplitudes are as follows
\begin{equation}
\Phi \left( {n,t} \right) = \left( {\begin{array}{*{20}{c}}
{\left\langle {{0,0,n}}
 \mathrel{\left | {\vphantom {{0,0,n} {{\psi _t}}}}
 \right. \kern-\nulldelimiterspace}
 {{{\psi _t}}} \right\rangle }\\
{\left\langle {{0,1,n}}
 \mathrel{\left | {\vphantom {{0,1,n} {{\psi _t}}}}
 \right. \kern-\nulldelimiterspace}
 {{{\psi _t}}} \right\rangle }\\
{\left\langle {{1,0,n}}
 \mathrel{\left | {\vphantom {{1,0,n} {{\psi _t}}}}
 \right. \kern-\nulldelimiterspace}
 {{{\psi _t}}} \right\rangle }\\
{\left\langle {{1,1,n}}
 \mathrel{\left | {\vphantom {{1,1,n} {{\psi _t}}}}
 \right. \kern-\nulldelimiterspace}
 {{{\psi _t}}} \right\rangle }
\end{array}} \right). \label{15}
\end{equation}
The shift operator is defined as in Eq. (\ref{6}). After one step of time evolution,
\begin{equation}
\Phi \left( {n,t + 1} \right) = \left( {\begin{array}{*{20}{c}}
{\left\langle {{0,0,n + 1}}
 \mathrel{\left | {\vphantom {{0,0,n + 1} {\left( {H \otimes {{\rm{I}}_2} \otimes {{\rm{I}}_N}} \right){\psi _t}}}}
 \right. \kern-\nulldelimiterspace}
 {{\left( {H \otimes {{\rm{I}}_2} \otimes {{\rm{I}}_N}} \right){\psi _t}}} \right\rangle }\\
{\left\langle {{0,1,n - 1}}
 \mathrel{\left | {\vphantom {{0,1,n - 1} {\left( {H \otimes {{\rm{I}}_2} \otimes {{\rm{I}}_N}} \right){\psi _t}}}}
 \right. \kern-\nulldelimiterspace}
 {{\left( {H \otimes {{\rm{I}}_2} \otimes {{\rm{I}}_N}} \right){\psi _t}}} \right\rangle }\\
{\left\langle {{1,1,n}}
 \mathrel{\left | {\vphantom {{1,1,n} {\left( {H \otimes {{\rm{I}}_2} \otimes {{\rm{I}}_N}} \right){\psi _t}}}}
 \right. \kern-\nulldelimiterspace}
 {{\left( {H \otimes {{\rm{I}}_2} \otimes {{\rm{I}}_N}} \right){\psi _t}}} \right\rangle }\\
{\left\langle {{1,0,n}}
 \mathrel{\left | {\vphantom {{1,0,n} {\left( {H \otimes {{\rm{I}}_2} \otimes {{\rm{I}}_N}} \right){\psi _t}}}}
 \right. \kern-\nulldelimiterspace}
 {{\left( {H \otimes {{\rm{I}}_2} \otimes {{\rm{I}}_N}} \right){\psi _t}}} \right\rangle }
\end{array}} \right). \label{16}
\end{equation}
Evaluating the action of the Hadamard matrix on the coin register, we can obtain
\[\begin{array}{l}
\Phi \left( {n,t + 1} \right) = \frac{1}{{\sqrt 2 }}\left( {\begin{array}{*{20}{c}}
{\left\langle {{0,0,n + 1}}
 \mathrel{\left | {\vphantom {{0,0,n + 1} {{\psi _t}}}}
 \right. \kern-\nulldelimiterspace}
 {{{\psi _t}}} \right\rangle  + \left\langle {{1,0,n + 1}}
 \mathrel{\left | {\vphantom {{1,0,n + 1} {{\psi _t}}}}
 \right. \kern-\nulldelimiterspace}
 {{{\psi _t}}} \right\rangle }\\
0\\
0\\
0
\end{array}} \right)\\
 + \frac{1}{{\sqrt 2 }}\left( {\begin{array}{*{20}{c}}
0\\
{\left\langle {{0,1,n - 1}}
 \mathrel{\left | {\vphantom {{0,1,n - 1} {{\psi _t}}}}
 \right. \kern-\nulldelimiterspace}
 {{{\psi _t}}} \right\rangle  + \left\langle {{1,1,n - 1}}
 \mathrel{\left | {\vphantom {{1,1,n - 1} {{\psi _t}}}}
 \right. \kern-\nulldelimiterspace}
 {{{\psi _t}}} \right\rangle }\\
0\\
0
\end{array}} \right) + \frac{1}{{\sqrt 2 }}\left( {\begin{array}{*{20}{c}}
0\\
0\\
{\left\langle {{0,1,n}}
 \mathrel{\left | {\vphantom {{0,1,n} {{\psi _t}}}}
 \right. \kern-\nulldelimiterspace}
 {{{\psi _t}}} \right\rangle  - \left\langle {{1,1,n}}
 \mathrel{\left | {\vphantom {{1,1,n} {{\psi _t}}}}
 \right. \kern-\nulldelimiterspace}
 {{{\psi _t}}} \right\rangle }\\
{\left\langle {{0,0,n}}
 \mathrel{\left | {\vphantom {{0,0,n} {{\psi _t}}}}
 \right. \kern-\nulldelimiterspace}
 {{{\psi _t}}} \right\rangle  - \left\langle {{1,0,n}}
 \mathrel{\left | {\vphantom {{1,0,n} {{\psi _t}}}}
 \right. \kern-\nulldelimiterspace}
 {{{\psi _t}}} \right\rangle }
\end{array}} \right).
\end{array}\]

Rewriting the above expression using $\Phi \left( {n + 1,t} \right)$, $\Phi \left( {n - 1,t} \right)$ and $\Phi \left( {n,t} \right)$, we can obtain
\begin{equation}
\Phi \left( {n,t + 1} \right) = {M_1}\Phi \left( {n + 1,t} \right) + {M_2}\Phi \left( {n - 1,t} \right) + {M_3}\Phi \left( {n,t} \right), \label{17}
\end{equation}
where ${M_1}$, ${M_2}$ and ${M_3}$ matrices are
\begin{equation}
{M_1} = \frac{1}{{\sqrt 2 }}\left( {\begin{array}{*{20}{c}}
1&0&1&0\\
0&0&0&0\\
0&0&0&0\\
0&0&0&0
\end{array}} \right),{M_2} = \frac{1}{{\sqrt 2 }}\left( {\begin{array}{*{20}{c}}
0&0&0&0\\
0&1&0&1\\
0&0&0&0\\
0&0&0&0
\end{array}} \right),{M_3} = \frac{1}{{\sqrt 2 }}\left( {\begin{array}{*{20}{c}}
0&0&0&0\\
0&0&0&0\\
0&1&0&{ - 1}\\
1&0&{ - 1}&0
\end{array}} \right). \label{18}
\end{equation}

In order to obtain the expression for the amplitudes of the quantum walk on the Cayley graph of the dihedral group, we use the method introduced in Ref. \cite{nayak2000quantum} and the analysis scheme given in Ref. \cite{Gettrick2014quan}. The time evolution of the walk using the Fourier transform is represented as follows
\begin{equation}
\begin{array}{l}
\tilde \Phi \left( {k,t + 1} \right) = \sum\limits_{n = 0}^{N - 1} {{e^{\frac{{2\pi ikn}}{N}}}\Phi \left( {n,t + 1} \right)} \\
 = \sum\limits_{n = 0}^{N - 1} {{e^{\frac{{2\pi ikn}}{N}}}\left( {{M_1}\Phi \left( {n + 1,t} \right) + {M_2}\Phi \left( {n - 1,t} \right) + {M_3}\Phi \left( {n,t} \right)} \right)} \\
 = {M_1}\sum\limits_n {{e^{\frac{{2\pi ikn}}{N}}}} \Phi \left( {n + 1,t} \right) + {M_2}\sum\limits_n {{e^{\frac{{2\pi ikn}}{N}}}} \Phi \left( {n - 1,t} \right) + {M_3}\sum\limits_n {{e^{\frac{{2\pi ikn}}{N}}}} \Phi \left( {n,t} \right)\\
 = \left( {{e^{ - \frac{{2\pi ik}}{N}}}{M_1} + {e^{\frac{{2\pi ik}}{N}}}{M_2} + {M_3}} \right)\tilde \Phi \left( {k,t} \right)
\end{array}. \label{19}
\end{equation}
From the above, we get a recursive relation for the time evolution in the Fourier basis
\begin{equation}
\tilde \Phi \left( {k,t} \right) = {\widetilde {{M_k}}^t}\tilde \Phi \left( {k,0} \right), \label{20}
\end{equation}
where the unitary matrix is as follows
\begin{equation}
{\widetilde M_k} = {e^{ - \frac{{2\pi ik}}{N}}}{M_1} + {e^{\frac{{2\pi ik}}{N}}}{M_2} + {M_3} = \frac{1}{{\sqrt 2 }}\left( {\begin{array}{*{20}{c}}
{{e^{ - \frac{{2\pi ik}}{N}}}}&0&{{e^{ - \frac{{2\pi ik}}{N}}}}&0\\
0&{{e^{\frac{{2\pi ik}}{N}}}}&0&{{e^{\frac{{2\pi ik}}{N}}}}\\
0&1&0&{ - 1}\\
1&0&{ - 1}&0
\end{array}} \right). \label{21}
\end{equation}

Here the characteristic polynomial of matrix $\widetilde {{M_k}}$ is the same as the characteristic polynomial of matrix $M_k$ in Ref. \cite{Gettrick2014quan}. Thus the eigenvalues are the same. Similarly, let us denote $A = w_N^{ - k} = {e^{ - \frac{{2\pi ik}}{N}}}$. We can also use symbolic variables and functions in MATLAB to find the eigenvalues and eigenvectors of $\widetilde {{M_k}}$. Matrix $\widetilde {{M_k}}$ has the following eigenvalues
\[\begin{array}{l}
{\lambda _1} =  - 1,{\lambda _3} = \frac{{1 + {A^2} + \sqrt {{A^4} - 6{A^2} + 1} }}{{2\sqrt 2 A}},\\
{\lambda _2} = 1,{\lambda _4} = \frac{{1 + {A^2} - \sqrt {{A^4} - 6{A^2} + 1} }}{{2\sqrt 2 A}},
\end{array}\]
with corresponding eigenvectors
\[\begin{array}{*{20}{l}}
{{v_1} = \left( {\begin{array}{*{20}{c}}
{ - \frac{A}{{\sqrt 2 A + 1}}}\\
{ - \frac{1}{{\sqrt 2 A + 1}}}\\
{\frac{{\sqrt 2 A + 2}}{{2A + \sqrt 2 }}}\\
1
\end{array}} \right),{v_3} = \left( {\begin{array}{*{20}{c}}
{\frac{{\sqrt {{A^4} - 6{A^2} + 1}  + {A^2} - 1}}{{2A}}}\\
{\frac{{{A^2} - \sqrt {{A^4} - 6{A^2} + 1}  - 1}}{{2{A^2}}}}\\
{ - \frac{1}{A}}\\
1
\end{array}} \right),}\\
{{v_2} = \left( {\begin{array}{*{20}{c}}
{\frac{A}{{\sqrt 2 A - 1}}}\\
{\frac{1}{{\sqrt 2 A - 1}}}\\
{ - \frac{{A - \sqrt 2 }}{{\sqrt 2 A - 1}}}\\
1
\end{array}} \right),{v_4} = \left( {\begin{array}{*{20}{c}}
{\frac{{{A^2} - \sqrt {{A^4} - 6{A^2} + 1}  - 1}}{{2A}}}\\
{\frac{{{A^2} + \sqrt {{A^4} - 6{A^2} + 1}  - 1}}{{2{A^2}}}}\\
{ - \frac{1}{A}}\\
1
\end{array}} \right).}
\end{array}\]
Thus, the spectral decomposition of $U$ can be given as follows
\begin{equation}
\begin{array}{l}
U = \sum\limits_{j = 0}^{N - 1} {\left( {{\lambda _1}\left| {{v_1},{\kappa _k}} \right\rangle \left. {\langle {v_1},{\kappa _k}} \right| + {\lambda _2}\left| {{v_2},{\kappa _k}} \right\rangle \left. {\langle {v_2},{\kappa _k}} \right|} \right.} \\
\left. { + {\lambda _3}\left| {{v_3},{\kappa _k}} \right\rangle \left. {\langle {v_3},{\kappa _k}} \right| + {\lambda _4}\left| {{v_4},{\kappa _k}} \right\rangle \left. {\langle {v_4},{\kappa _k}} \right|} \right),
\end{array}  \label{22}
\end{equation}
where $\left| {{\kappa _k}} \right\rangle  = \frac{1}{{\sqrt N }}\sum\limits_{j = 0}^{N - 1} {w_N^{jk}\left| j \right\rangle }$ is the spatial part of the computational basis on which the Fourier transform acts and ${w_N} = {e^{\frac{{2\pi i}}{N}}}$ is a primitive ${N^{th}}$ root of unity.

\section{Relation with Quantum Walks with Memory on Cycles}

In this part, we give deeper thought to the relation between quantum walk on the Cayley graph of the dihedral group and quantum walk with memory on a cycle from different angels. We explore the walk model based on the characteristics of group members on the one hand, and on the other hand, we briefly discuss the one-to-one correspondence between quantum walk with memory on the line digraph of the cycle and quantum walk without memory on the Cayley graph of the dihedral group.

\subsection{From Hilbert Space and Eigenvalues Perspective}

In Ref. \cite{Gettrick2014quan}, Mc Gettrick introduced a model of quantum walk with memory on a cycle and studied its basic properties. We find that the walk operators of quantum walk without memory on the Cayley graph of the dihedral group $D_N$ and quantum walk with one-step memory on an $N$-cycle have the same eigenvalues. The quantum walk with one-step memory on an $N$-cycle is defined in Hilbert space $H_C^2 \otimes H_M^2 \otimes H_V^N$ and the quantum walk without memory on the Cayley graph of the dihedral group is defined in Hilbert space $H_C^2 \otimes H_V^{2N} = H_C^2 \otimes H_V^2 \otimes H_V^N$. They are equivalent. The eigenvectors which correspond to the same eigenvalues are different and just affect computational basis state. Furthermore, the basic properties of the two walks associated with eigenvalues and eigenvectors are homologous. The two main properties are probability distribution and time-averaged limiting probability distribution. Probability distribution and the time-averaged value of the walk are referred to Ref. \cite{Gettrick2014quan} and we make a simple description of them in the Appendix.

We evaluate the probability distribution of finding the particle at each vertex of the Cayley graph of the dihedral group. The probability distribution of the walk without memory after the first $n$ steps is given by
\begin{equation}
{p_{sj}}\left( n \right) = \sum\limits_{c} {{{\left| {\left\langle {{c,s,j}}
 \mathrel{\left | {\vphantom {{c,s,j} {{\psi _n}}}}
 \right. \kern-\nulldelimiterspace}
 {{{\psi _n}}} \right\rangle } \right|}^2}}. \label{23}
\end{equation}

In Hilbert space $H_C^2 \otimes H_V^2 \otimes H_V^N$, we first consider part of Hilbert space, i.e. $H_C^2 \otimes H_V^2$. If the walker starts from the vertex $\left( {0,0} \right)$ with the coin in the state ${\left| 0 \right\rangle }$, the initial state is as follows
\begin{equation}
\left| {{\psi _0}} \right\rangle  = {\left| 0 \right\rangle _C} \otimes {\left| 0 \right\rangle _V}{\left| 0 \right\rangle _V} = \left( {\begin{array}{*{20}{c}}
1\\
0\\
0\\
0
\end{array}} \right) \otimes {\left| 0 \right\rangle _V}. \label{24}
\end{equation}
If the starting vertex is $\left( {1,0} \right)$, the initial state is
\begin{equation}
\left| {{\psi _0}} \right\rangle  = {\left| 0 \right\rangle _C} \otimes {\left| 1 \right\rangle _V}{\left| 0 \right\rangle _V} = \left( {\begin{array}{*{20}{c}}
0\\
1\\
0\\
0
\end{array}} \right) \otimes {\left| 0 \right\rangle _V}. \label{25}
\end{equation}
For the quantum walk with one-step memory on an $N$-cycle, the memory register is in the superposition $\frac{1}{{\sqrt 2 }}\left( {\left| 0 \right\rangle  + \left| 1 \right\rangle } \right)$, the coin register is in the state ${\left| 0 \right\rangle }$ and the starting position is $0$, which means that the particle of the quantum walk on the Cayley graph of the dihedral group is starting from the vertex $\left( {0,0} \right)$ and the vertex $\left( {1,0} \right)$ at the same time, i.e.
\begin{equation}
\left| {{\psi _0}} \right\rangle  = \left( {\begin{array}{*{20}{c}}
1\\
1\\
0\\
0
\end{array}} \right) \otimes {\left| 0 \right\rangle _V}. \label{26}
\end{equation}
The formula for probability distribution is given in Ref. \cite{Gettrick2014quan} and it is introduced by Eqs. (\ref{app:1}-\ref{app:6}) in the Appendix.

In Ref. \cite{aharonov2001quantum}, it was shown that probability distribution is quasi-periodic for quantum walks on cycles and it was suggested to consider time-averaged probability distribution, which converges to the limiting distribution with $t \to \infty $. Here, we focus on the probability distribution of quantum walk on the Cayley graph of the dihedral group. For a fixed vertex $\left( {s,j} \right)$, the probability ${p_{sj}}\left( n \right)$ does not converge to a limit. The time-averaged probability ${\bar p_{sj}}\left( n \right)$ is considered instead of ${{p_{sj}}\left( n \right)}$ and the probability distribution of the walk without memory after the first $n$ steps is given by
\begin{equation}
{{\bar p}_{sj}}\left( n \right) = \frac{1}{n}\sum\limits_{i = 1}^n {{p_{sj}}\left( i \right)}.\label{27}
\end{equation}
The formula for time-averaged limiting probability distribution is given by Eqs. (\ref{app:5}) and (\ref{app:6}). It is worth reminding that there is a prerequisite that the particle of the quantum walk on the Cayley graph of the dihedral group starts from the vertex $\left( {0,j} \right)$ and the vertex $\left( {1,j} \right)$ at the same time, where $0 \le j \le N - 1$.

\subsection{From Correspondence Perspective}

In Ref. \cite{Generic2016}, Dan Li et al. presented the one-to-one correspondence between quantum walks with memory(QWM) on a regular graph and quantum walks without memory(QWoM) on line digraph of the regular graph. And inspired by Ref. \cite{graphwalks2012}, they constructed the model of generic QWoM on line digraph of an $m$-regular graph.

Here we introduce two definitions to explain the relationship between the two walks, i.e. QWM on cycles and QWoM on the Cayley graph of the dihedral group.

\begin{definition}
\textbf{(Line digraph)}
Let $G(V,E)$ be a digraph with vertex set $V(G)$ and edge set $E(G)$. Define $\overrightarrow{L}G$ be the line digraph of a digraph $G(V,E)$. The vertex set of $\overrightarrow{L}G$ is $E(G)$; for ${x_a},{x_b},{x_c},{x_d} \in V\left( G \right),\left( {\left( {{x_a},{x_b}} \right),\left( {{x_c},{x_d}} \right)} \right) \in E\left( {\overrightarrow{L}G} \right)$ if and only if ${\left( {{x_a},{x_b}} \right)}$ and ${\left( {{x_c},{x_d}} \right)}$ are both in $E(G)$ and ${{x_b} = {x_c}}$. The line digraph of $\overrightarrow{L}G$ is denoted by $\overrightarrow{L}^{2}G$. Similarity, there are $\overrightarrow{L}^{d}G$ with $d \in {N^ + }$ and all of them is called as line digraph of $G(V,E)$.
\end{definition}

\begin{definition}
\textbf{(Partition)}
Let $G(V,E)$ an $m$-regular graph. Define $\pi $ be a partition of $\overrightarrow{L}^{d}G$ such that
\[\pi :\overrightarrow{L}^{d}G \to \left\{ {{C_1},{C_2}, \cdots ,{C_m}} \right\},\]
where $\left\{ {{C_k}|k = 1, \cdots ,m} \right\}$ satisfy that $V\left( {{C_k}} \right) = V\left( {\overrightarrow{L}^{d}G} \right)$, ${ \cup _k}E\left( {{C_k}} \right) = E\left( {\overrightarrow{L}^{d}G} \right)$ and for every vertex $v \in V\left( {{C_k}} \right)$, the outdegree is 1. Directed cycle factorization is a kind of partition which satisfies that for every vertex $v \in V\left( {{C_k}} \right)$, the outdegree and indegree are 1. We denote the set of partitions of $\overrightarrow{L}^{d}G$ by ${\Pi _{\overrightarrow{L}^{d}G}}$.
\end{definition}

\begin{figure}
\centering
\subfigure[]{
\includegraphics[height=3.4cm]{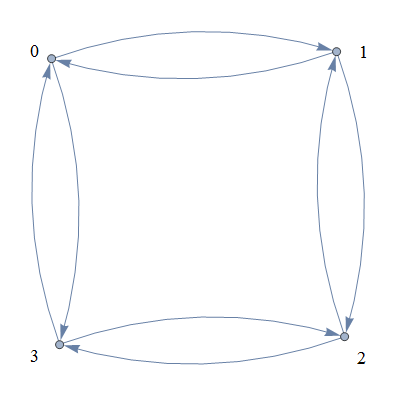}
}
\subfigure[]{
\includegraphics[height=3.4cm]{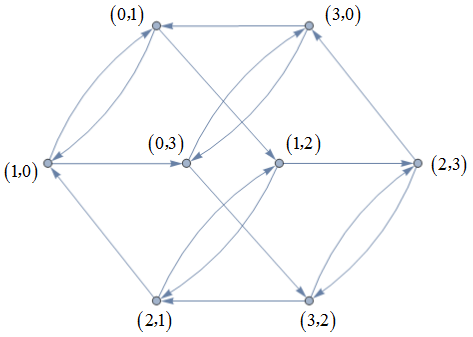}
}
\subfigure[]{
\includegraphics[height=3.4cm]{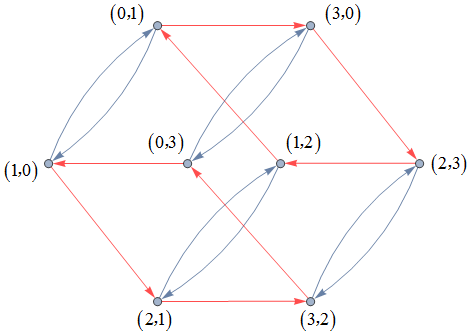}
}
\subfigure[]{
\includegraphics[height=3.4cm]{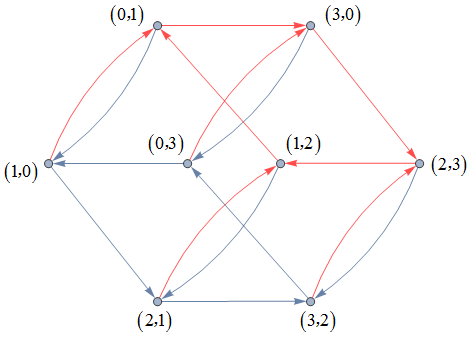}
}
\caption{The original graph $G(V,E)$ and the line digraph of $G(V,E)$. Subgraph (c) and (d) show two partitions of the line digraph of $G(V,E)$ by using different color lines to denote $C_k$.}
\label{fig:3}
\end{figure}

Using the notation in the definition above, the generic QWoM on ${\overrightarrow{L}^{d}G}$ means generic QWM with $d$-step memory on an $m$-regular graph $G(V,E)$. In this part, we focus on QWM with one-step memory on cycles, i.e. $m = 2,d = 1$. We show an example of the line digraph in Fig.\ref{fig:3}. The original graph $G(V,E)$ is a $4$-cycle in Fig.\ref{fig:3}(a). The line digraph of $G(V,E)$, i.e. ${\overrightarrow{L}G}$, is shown in Fig.\ref{fig:3}(b). We show two partition{color{red}s} ${\pi _1}$ and ${\pi _2}$ in Fig.\ref{fig:3}(c, d), respectively.

From the perspective of graph theory, Fig.\ref{fig:3}(b) and Fig.\ref{fig:1}(a) are isomorphic. Similarly, the line digraph of an $N$-cycle, denoted by ${\overrightarrow{L}G}$, and the Cayley graph of dihedral group $D_N$ are isomorphic. From this perspective, QWoM on the Cayley graph of the dihedral group $D_N$ and QWM with one-step memory on an $N$-cycle are equivalent.

\section{Numerical Simulation}

In this part, numerical simulation of the quantum walk on the Cayley graph of the dihedral group is provided with MATLAB code to study the shift matrix and the walk matrix. The numerical simulation helps us verify the results of theoretical analysis. We focus on the Hadamard walk on the Cayley graph of the dihedral group. It is worth mentioning that we change the order of tensor product operation. Hilbert space of the quantum walk on the Cayley graph of the dihedral group is $H_V^{2N} \otimes H_C^2 = H_V^2 \otimes H_V^N \otimes H_C^2$ and the shift operator is as follows
\[\begin{array}{*{20}{l}}
{\hat S\left| 0 \right\rangle \left| j \right\rangle {{\left| R \right\rangle }_C} = \left| 0 \right\rangle \left| {j + 1} \right\rangle {{\left| R \right\rangle }_C}}\\
{\hat S\left| 0 \right\rangle \left| j \right\rangle {{\left| F \right\rangle }_C} = \left| 1 \right\rangle \left| j \right\rangle {{\left| F \right\rangle }_C}}\\
{\hat S\left| 1 \right\rangle \left| j \right\rangle {{\left| R \right\rangle }_C} = \left| 1 \right\rangle \left| {j - 1} \right\rangle {{\left| R \right\rangle }_C}}\\
{\hat S\left| 1 \right\rangle \left| j \right\rangle {{\left| F \right\rangle }_C} = \left| 0 \right\rangle \left| j \right\rangle {{\left| F \right\rangle }_C}}\\
{j \in {Z_N}}
\end{array}.\]
The evolution operator is $\hat U = \hat S \cdot \left( {\hat I_V \otimes \hat C} \right)$.

Through numerical simulation, we can get the shift matrix $S$. Define an $N$-dimensional vector, $basis\left( {N,i} \right)$, as the $i$th vector of the standard basis. The shift matrix $S$ can be represented as
\[S = \left( {\begin{array}{*{20}{c}}
A&B\\
B&C
\end{array}} \right),\]
where block matrix $A$, $B$ and $C$ are all $N$-dimensional. According to the parity of the column index, matrix elements are as follows
\begin{equation}
A(:,i) =
\begin{cases}
basis(N,(i+2) mod N) & \text{if $i$ is odd} \\
\bf{0} & \text{if $i$ is even}
\end{cases}
\end{equation}
\begin{equation}
B(:,i) =
\begin{cases}
\bf{0} & \text{if $i$ is odd} \\
basis(N,(i+2) mod N) & \text{if $i$ is even}
\end{cases}
\end{equation}
\begin{equation}
C(:,i) =
\begin{cases}
basis(N,(i-2) mod N) & \text{if $i$ is odd} \\
\bf{0} & \text{if $i$ is even}
\end{cases}
\end{equation}
According to Eqs. (\ref{13}) and (\ref{14}), we can get the walk matrix $U$. The size of both of shift matrix $S$ and the walk matrix $U$ are $4N \times 4N$. Naturally, we can verify the theoretical analysis results and begin further research using numerical simulation.

\begin{figure}
\centering
\subfigure[]{
\includegraphics[height=3.5cm]{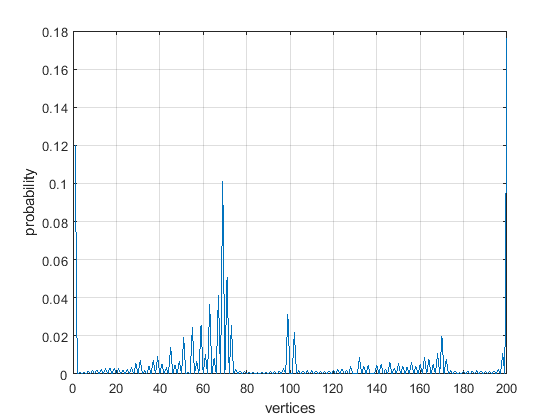}
}
\subfigure[]{
\includegraphics[height=3.5cm]{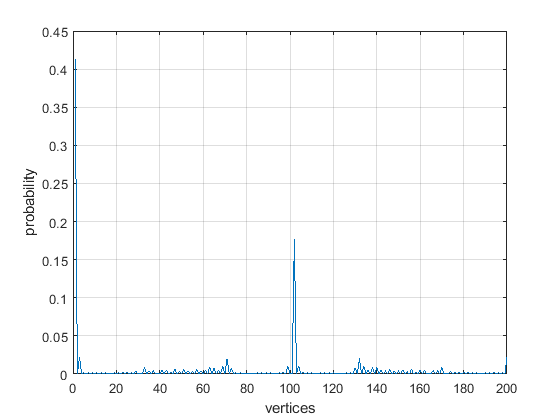}
}
\subfigure[]{
\includegraphics[height=3.5cm]{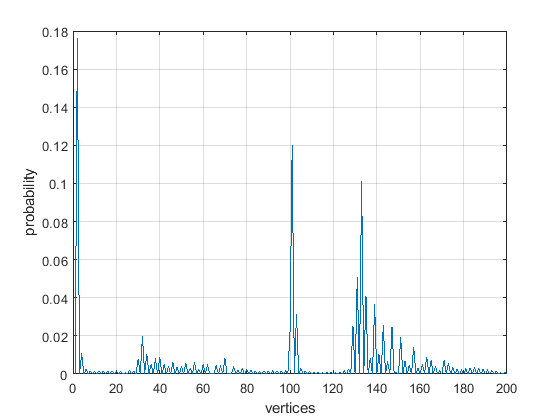}
}
\subfigure[]{
\includegraphics[height=3.5cm]{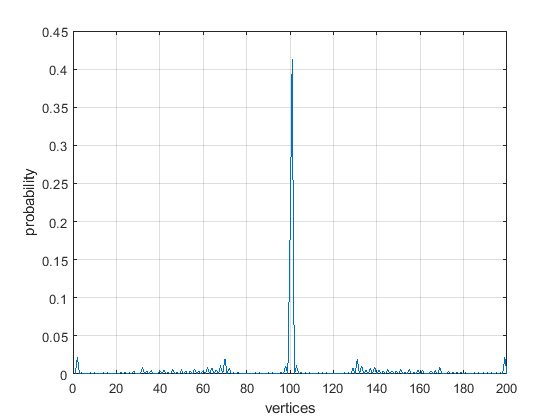}
}
\caption{Probability distribution of the quantum walk on the Cayley graph of the dihedral group $D_N$ with $N = 100$ after 100 steps using the coin and shift operators given by Eqs. (\ref{3}) and (\ref{6}), respectively. Plot (a) corresponds to a discrete quantum walk on the Cayley graph of the dihedral group with total initial state ${\left| 0 \right\rangle _V}{\left| 0 \right\rangle _V}{\left| 0 \right\rangle _C}$, while plot (b) has total initial state ${\left| 0 \right\rangle _V}{\left| 0 \right\rangle _V}{\left| 1 \right\rangle _C}$. Plot (c) corresponds to a discrete quantum walk on the Cayley graph of the dihedral group with total initial state ${\left| 1 \right\rangle _V}{\left| 0 \right\rangle _V}{\left| 0 \right\rangle _C}$, while plot (d) has total initial state ${\left| 1 \right\rangle _V}{\left| 0 \right\rangle _V}{\left| 1 \right\rangle _C}$.}
\label{fig:4}
\end{figure}

In order to evaluate the probability distribution, we need to study how the initial state affects the property of the walk. If the coin operator is the Hadamard operator, we take four initial states into account, i.e.
\[\begin{array}{*{20}{l}}
{\left| {\psi \left( 0 \right)} \right\rangle  = {{\left| 0 \right\rangle }_C} \otimes {{\left| 0 \right\rangle }_V}{{\left| 0 \right\rangle }_V},\left| {\psi \left( 0 \right)} \right\rangle  = {{\left| 0 \right\rangle }_C} \otimes {{\left| 1 \right\rangle }_V}{{\left| 0 \right\rangle }_V},}\\
{\left| {\psi \left( 0 \right)} \right\rangle  = {{\left| 1 \right\rangle }_C} \otimes {{\left| 0 \right\rangle }_V}{{\left| 0 \right\rangle }_V},\left| {\psi \left( 0 \right)} \right\rangle  = {{\left| 1 \right\rangle }_C} \otimes {{\left| 1 \right\rangle }_V}{{\left| 0 \right\rangle }_V},}
\end{array}\]
where the coin state is ${\left| 0 \right\rangle _C}$ or ${\left| 1 \right\rangle _C}$ in Hilbert space $H_C^2$, and the initial vertex state is ${\left| 0 \right\rangle _V}{\left| 0 \right\rangle _V}$ or ${\left| 1 \right\rangle _V}{\left| 0 \right\rangle _V}$ in Hilbert space $H_V^{2N} = H_V^2 \otimes H_V^N$.
The shift operator allows the walker to go one step rotation if the accompanying coin state is ${\left| 0 \right\rangle _C}$, or one step flip if the accompanying coin state is the other basis state ${\left| 1 \right\rangle _C}$. ${\left| 0 \right\rangle _V}{\left| 0 \right\rangle _V}$ and ${\left| 1 \right\rangle _V}{\left| 0 \right\rangle _V}$ represent the rotation and the flip element of the dihedral group, respectively. The numerical simulation results of probability distribution of the quantum walk on the Cayley graph of the dihedral group $D_N$ with $N = 100$ after 100 steps are shown in Fig.\ref{fig:4}.

According to Fig.\ref{fig:4}(a, c) and Fig.\ref{fig:4}(b, d), the probability distribution always fluctuates wildly. We can also study the property of the quantum walk on the Cayley graph of the dihedral group and compare the properties of the quantum walk on the Cayley graph of the dihedral group and on the line, or on a cycle. If the initial coin state is $\frac{1}{{\sqrt 2 }}\left( {{{\left| 0 \right\rangle }_C} + i{{\left| 1 \right\rangle }_C}} \right)$, we can not obtain a symmetric probability distribution.

\begin{figure}
\centering
\subfigure[]{
\includegraphics[height=5.4cm]{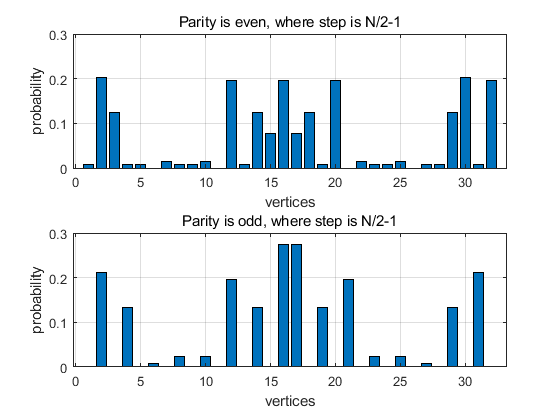}
}
\subfigure[]{
\includegraphics[height=5.4cm]{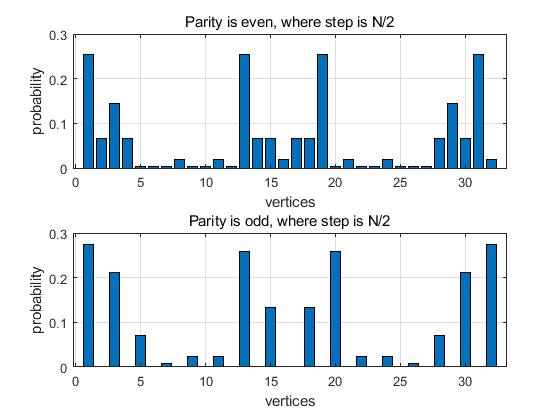}
}
\subfigure[]{
\includegraphics[height=5.4cm]{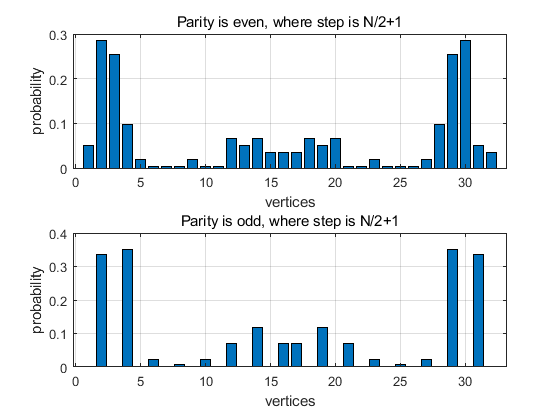}
}
\caption{The probability of finding the particle at vertex $\left( {s,j} \right)$ at the (a) ${{\frac{N}{2}-1}^{th}}$, (b) ${\frac{N}{2}^{th}}$ step and (c) ${{\frac{N}{2}+1}^{th}}$, where $N = {2^4}$ and $j$ is random in $Z_N$.}
\label{fig:5}
\end{figure}

We study the probability distribution of the quantum walk on the Cayley graph of the dihedral group starting from the initial state
\begin{equation}
{\left| {{\psi _0}} \right\rangle  = {{\left| 0 \right\rangle }_C} \otimes \left( {{{\left| 0 \right\rangle }_V}{{\left| j \right\rangle }_V} + {{\left| 1 \right\rangle }_V}{{\left| {j + d} \right\rangle }_V}} \right)}.
\end{equation}
By using numerical simulation method, we can observe the probability of finding the particle at vertex $\left( {s,j} \right)$ after $n$ steps. We find that the parity of $d$ affects probability distribution of the quantum walk on the Cayley graph of the dihedral group. A specific example is illustrated in Fig.\ref{fig:5}, where $N = {2^4}$ and $j$ is random in $Z_N$. We explain explicitly that if $d$ is even, then the state is $\left| {{\psi _{\frac{N}{2}}}} \right\rangle  = \sum\limits_{c = 0}^1 {\sum\limits_{s = 0}^1 {\sum\limits_{j = 0}^{N - 1} {{\alpha _{csj}}\left| {c,s,j} \right\rangle } } } $ and amplitudes ${{\alpha _{csj}}}$ are all non-zero. If $d$ is odd, half of all amplitudes ${{\alpha _{csj}}}$ are 0. It is worth reminding that the bigger $n$ is, where $N = {2^n}$, the lower the probability is. In this case, some amplitudes ${{\alpha _{csj}}}$ are infinitesimally small-probability.

\begin{figure}
\centering
\subfigure[]{
\includegraphics[height=4.3cm]{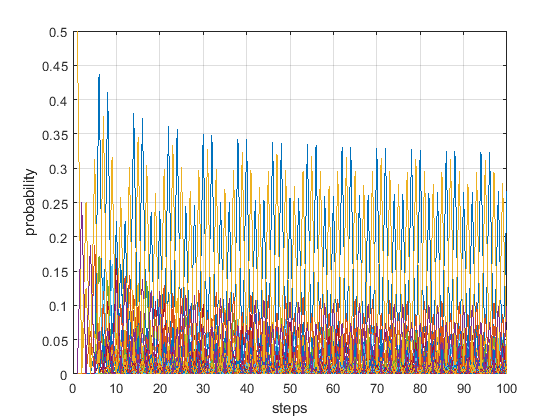}
}
\subfigure[]{
\includegraphics[height=4.3cm]{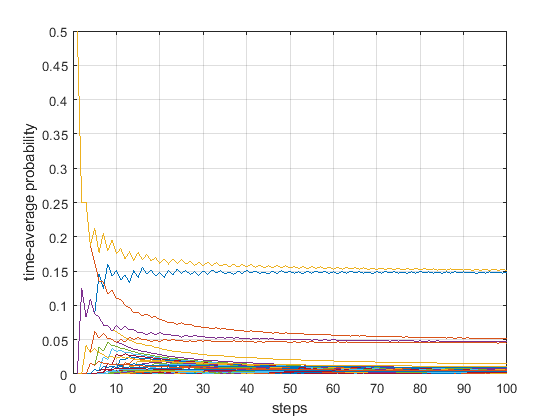}
}
\caption{Probability distribution and time-averaged probability distribution of quantum walk without memory on the Cayley graph of the dihedral group $D_N$ with $N = 100$ after the first 100 steps using the coin and shift operators given by Eqs. (\ref{3}) and (\ref{6}) respectively. Plot (a) corresponds to Eq. (\ref{23}), while plot (b) corresponds to Eq. (\ref{27}), with the initial coin state $\frac{1}{{\sqrt 2 }}\left( {{{\left| 0 \right\rangle }_C} + i{{\left| 1 \right\rangle }_C}} \right)$. }
\label{fig:6}
\end{figure}

The numerical simulation graphs of the probability distribution and the time-averaged probability distribution for a fixed vertex of the Cayley graph of the dihedral group are shown in Fig.\ref{fig:6}. It can be seen that probability distribution of the quantum walk on the Cayley graph of the dihedral group does not converge, while time-averaged probability distribution converges with $t \to \infty $ to the limiting distribution.

\section{Summary}

In this paper, we have introduced a model of discrete-time quantum walk on the Cayley graph of the dihedral group. The model we proposed opens the door to quantum walks on non-abelian groups. The Abelian group is to cyclic quantum walk, what the dihedral group is to the present ``dihedral quantum  walk", i.e. the non-abelian case. We also study the relation between quantum walk on the Cayley graph of the dihedral group and quantum walk with memory on a cycle. This exploration is truly novel. It appears that the additional generator $\tau$ (for generating flips) in the dihedral case corresponds to memory in the cycles. Thus, the physical implication for quantum walks of going from Abelian group (rotational symmetry of the regular $N$-gon) to the dihedral group (full symmetry of the regular $N$-gon) maybe motivate further research on QWoM and QWM, thereby giving a nice physical representation to an abstract mathematical concept. Moreover, we provide a numerical simulation of the quantum walk on the Cayley graph of the dihedral group in MATLAB and the simulation results validate the correctness of the theory and show the validity of the method.

Our work extends current theoretical researches on quantum walks and offers a new direction to study QWoM and QWM. In the future, we will consider quantum walk with memory on the Cayley graph of the dihedral group and compare the similarities and differences between quantum walks without and with memory on the Cayley graph of the dihedral group. Furthermore, we will consider whether the research on quantum walks without and with memory on the Cayley graph of the dihedral group can help get an efficient solution to Dihedral Hidden Subgroup Problem.

\subsubsection*{Acknowledgments.}This work was supported by the National Natural Science Foundation of China (Grant Nos. 61571226, 61701229, 61572053, 61702367), Natural Science Foundation of Jiangsu Province, China (Grant No. BK20170802), Postdoctoral Science Foundation funded Project of China (Grant Nos. 2018M630557, 2018T110499), Jiangsu Planned Projects for Postdoctoral Research Funds (Grant No. 1701139B), the Beijing Natural Science Foundation (Grant No. 4162005), the Research Project of Tianjin Municipal Commission of Education(Grant No. 2017KJ033).

\section*{Appendix: Probability Distribution of Quantum walk with memory on a cycle}

Here, we make a simple description of probability distribution and the time-averaged value of the walk referred to Ref. \cite{Gettrick2014quan}.

In quantum walk with memory on a cycle, the initial state of the walk is $\left| {{\phi _0}} \right\rangle $ and the vector of amplitudes is $\Phi \left( {n,0} \right)$. In the Fourier basis, the initial state is $\tilde \Phi \left( {k,0} \right)$, for any $k = 0, \cdots ,d - 1$, where $d$ is the number of vertices. And the form of the amplitudes after $t$ steps can be calculated and the vector of the amplitudes is
\begin{equation}
\tilde \Phi \left( {k,t} \right) = M_k^t\tilde \Phi \left( {k,0} \right), \label{app:1}
\end{equation}
for any $k$. The initial state of the walk $\tilde \Phi \left( {k,0} \right)$ can be written as $\tilde \Phi \left( {k,0} \right) = \sum\limits_{i = 1}^4 {{\alpha _i}\left( k \right){v_i}\left( k \right)} $ in the ${\left\{ {{v_i}} \right\}_{i = 1, \cdots ,4}}$ basis, where ${\alpha _i}\left( k \right) = \left( {{v_i}\left( k \right),\tilde \Phi \left( {k,0} \right)} \right)$ are components of ${\tilde \Phi \left( {k,0} \right)}$ in the ${\left\{ {{v_i}} \right\}_{i = 1, \cdots ,4}}$ basis. Thus the evolution in the Fourier basis can be written as
\begin{equation}
\tilde \Phi \left( {k,t} \right) = M_k^t\sum\limits_{i = 1}^4 {{\alpha _i}\left( k \right){v_i}\left( k \right)}  = \sum\limits_{i = 1}^4 {{\alpha _i}\left( k \right)\lambda _i^t\left( k \right){v_i}\left( k \right)}, \label{app:2}
\end{equation}
where ${\lambda _i}$ and ${v_i}$ are the eigenvalues and eigenvectors, respectively.

The original components can be expressed by the Fourier-transformed vectors as
\begin{equation}
\Phi \left( {n,t} \right) = \frac{1}{d}\sum\limits_{k = 0}^{d - 1} {{e^{\frac{{2\pi ikn}}{d}}}\tilde \Phi \left( {k,t} \right)}  = \frac{1}{d}\sum\limits_{k = 0}^{d - 1} {\sum\limits_{j = 1}^4 {{e^{\frac{{2\pi ikn}}{d}}}{\alpha _j}\left( k \right)\lambda _j^t\left( k \right){v_j}\left( k \right)} }. \label{app:3}
\end{equation}
Therefore the probability of finding the particle at the $n$th node after $t$ steps is
\begin{equation}
\begin{array}{l}
p\left( {n,t} \right) = {\left| {\Phi \left( {n,t} \right)} \right|^2}\\
 = \frac{1}{{{d^2}}}\sum\limits_{k,m = 0}^{d - 1} {\sum\limits_{j,l = 1}^4 {{e^{\frac{{2\pi i\left( {m - k} \right)n}}{d}}}\alpha _j^*\left( k \right){\alpha _l}\left( m \right)v_j^\dag \left( k \right){v_l}\left( m \right){{\left[ {\lambda _j^*\left( k \right){\lambda _l}\left( m \right)} \right]}^t}} } .
\end{array} \label{app:4}
\end{equation}

Time-averaged probability distribution is
\begin{equation}
\begin{array}{l}
\bar p\left( n \right) = \mathop {\lim }\limits_{t \to \infty } \frac{1}{t}\sum\limits_{s = 0}^{t - 1} {p\left( {n,s} \right)} \\
 = \frac{1}{{{d^2}}}\sum\limits_{k,m = 0}^{d - 1} {\sum\limits_{j,l = 1}^4 {{e^{\frac{{2\pi i\left( {m - k} \right)n}}{d}}}\alpha _j^*\left( k \right){\alpha _l}\left( m \right)v_j^\dag \left( k \right){v_l}\left( m \right)\mathop {\lim }\limits_{t \to \infty } \frac{1}{t}\sum\limits_{s = 0}^{t - 1} {{{\left[ {\lambda _j^*\left( k \right){\lambda _l}\left( m \right)} \right]}^s}} } } .
\end{array} \label{app:5}
\end{equation}
One can observe that the convergence of $\bar p\left( n \right)$ depends only on the behavior of the term
\begin{equation}
{\mathop {\lim }\limits_{t \to \infty } \frac{1}{t}\sum\limits_{s = 0}^{t - 1} {{{\left[ {\lambda _j^*\left( k \right){\lambda _l}\left( m \right)} \right]}^s}} }. \label{app:6}
\end{equation}
If ${\lambda _j^*\left( k \right){\lambda _l}\left( m \right) = 1}$, $\mathop {\lim }\limits_{t \to \infty } \frac{1}{t}\sum\limits_{s = 0}^{t - 1} {{{\left[ {\lambda _j^*\left( k \right){\lambda _l}\left( m \right)} \right]}^s}}  = 1$; otherwise 0. Unfortunately, any further simplifications of Eq. (\ref{app:4}) and (\ref{app:5}) were not possible. However, definition of $\mathop {\lim }\limits_{t \to \infty } \frac{1}{t}\sum\limits_{s = 0}^{t - 1} {{{\left[ {\lambda _j^*\left( k \right){\lambda _l}\left( m \right)} \right]}^s}} $ can be easily calculated using the standard computer algebra systems, and thus allows for the evaluation of Eq. (\ref{app:5}).
\end{document}